\documentstyle[aasms4]{article}
\makeatletter
  
  \@addtoreset{equation}{section}
\makeatother

\begin{document}

\title{RELATIVISTIC CORRECTIONS TO THE SUNYAEV-ZEL'DOVICH EFFECT FOR CLUSTERS OF GALAXIES. II.  INCLUSION OF PECULIAR VELOCITIES}

\author{SATOSHI NOZAWA\altaffilmark{1}}

\affil{Josai Junior College for Women, 1-1 Keyakidai, Sakado-shi, Saitama, 350-0295, Japan}

\author{NAOKI ITOH\altaffilmark{2}}

\affil{Department of Physics, Sophia University, 7-1 Kioi-cho, Chiyoda-ku, Tokyo, 102-8554, Japan}

\centerline{AND}

\author{YASUHARU KOHYAMA\altaffilmark{3}}

\affil{Fuji Research Institute Corporation, 2-3 Kanda-Nishiki-cho, Chiyoda-ku, Tokyo, 101-8443, Japan}

\altaffiltext{1}{snozawa@venus.josai.ac.jp}
\altaffiltext{2}{n\_itoh@hoffman.cc.sophia.ac.jp}
\altaffiltext{3}{kohyama@crab.fuji-ric.co.jp}

\begin{abstract}

  We extend the formalism of the relativistic thermal Sunyaev-Zel'dovich effect to the system moving with a velocity $\vec{\beta} \equiv \vec{v}/c$ with respect to the cosmic microwave background radiation.  In the present formalism, the kinematic Sunyaev-Zel'dovich effect for the cluster of galaxies with a peculiar velocity $\vec{\beta}$ is derived in a straightforward manner by the Lorentz boost of the generalized Kompaneets equation.  We give an analytic expression for the kinematic Sunyaev-Zel'dovich effect which is valid up to O($\beta^{2}$) with the power series expansion approximation in terms of $\theta_{e} \equiv k_{B} T_{e}/mc^{2}$, where $T_{e}$ and $m$ are the electron temperature and the electron mass, respectively.  It is found that the relativistic corrections to the kinematic Sunyaev-Zel'dovich effect are significant.  For a typical electron temperature $k_{B} T_{e}$ = 10keV, one obtains $-8.2\%$ and $+1.3\%$ corrections from the $O(\beta \theta_{e})$ and $O(\beta \theta_{e}^{2})$ contributions, respectively.  The $O(\beta^{2})$ correction is extremely small, $+0.2\%$ for $\beta=1/300$ at $k_{B} T_{e}$ = 10keV.  Therefore it can be safely neglected.  These relativistic corrections are directly reflected on the determination of the peculiar velocity $\beta$ of the cluster of galaxies with the observation of the kinematic Sunyaev-Zel'dovich effect. 

\end{abstract}


\section{INTRODUCTION}

  Compton scattering of the cosmic microwave background radiation (CMBR) by hot intracluster gas --- the thermal Sunyaev-Zel'dovich effect --- (Zel'dovich \& Sunyaev 1969; Sunyaev \& Zel'dovich 1972; Sunyaev \& Zel'dovich 1980a) provides a useful method to measure the Hubble constant $H_{0}$ (Gunn 1978; Silk \& White 1978; Birkinshaw 1979; Cavaliere, Danese, \& De Zotti 1979; Birkinshaw, Hughes, \& Arnaud 1991; Birkinshaw \& Hughes 1994; Myers et al. 1995; Herbig et al. 1995; Jones 1995; Markevitch et al. 1996; Holzapfel et al. 1997a; Holzapfel et al. 1997b).  The Sunyaev-Zel'dovich formula has been derived from a kinetic equation for the photon distribution function taking into account the Compton scattering by electrons --- the Kompaneets equation --- (Kompaneets 1957; Weymann 1965).  The Kompaneets equation has been derived with the non-relativistic approximation for the electron.  However, the electrons in the clusters of galaxies are extremely hot, $k_{B} T_{e}$ = $5 \sim 15$keV (Arnaud et al. 1994; Markevitch et al. 1994; Markevitch et al. 1996; Holzapfel et al. 1997a; Holzapfel et al. 1997b).

  Recently attempts have been made to include the relativistic corrections in the Sunyaev-Zel'dovich effect (Rephaeli 1995; Rephaeli \& Yankovitch 1997).  However, it appears that the calculations have not been carried out in a manifestly covariant form.  For example, equation (4) in Rephaeli (1995), which comes from Chandrasekhar (1950), is a non-relativistic formula.  Since the extension of the Kompaneets equation to the relativistic regime is extremely important in view of many recent measurements of the Hubble constant $H_{0}$ with the use of the Sunyaev-Zel'dovich effect, we will solve the kinetic equation for the photon distribution function in a manifestly covariant form taking into account the Compton scattering by electrons.

  Very recently a generalized Kompaneets equation has been derived by several groups: Stebbins (1997); Challinor \& Lasenby (1998); Itoh, Kohyama \& Nozawa (1998).  By using the generalized Kompaneets equation, analytic expressions for the Sunyaev-Zel'dovich effect have been derived as a power series of $\theta_{e}$ = $k_{B}T_{e}/mc^{2}$, where $T_{e}$ and $m$ are the electron temperature and the electron mass, respectively.  It has been shown that the results obtained by the power series expansion agree with the previous numerical calculations by Rephaeli (1995) and Rephaeli \& Yankovitch (1997), thereby proving the validity of their method.  It is really impressive to see their result agreeing with the result of the manifestly covariant calculation.  In particular, the convergence of the power series expansion has been carefully studied in the paper by Itoh, Kohyama \& Nozawa (1998), where the analytic expressions up to $O(\theta_{e}^{5})$ have been derived and the results have been compared with those of the direct numerical integration of the Boltzmann equation.  It has been shown that the power series expansion approximation is sufficiently accurate for the region $k_{B}T_{e} \leq 15$keV by taking into account up to $O(\theta_{e}^{5})$ contributions.

  On the other hand, Sunyaev \& Zel'dovich (1980b) have predicted that the CMBR spectrum is also distorted if the cluster of galaxies has a peculiar velocity with respect to the CMBR.  This effect is called the kinematic Sunyaev-Zel'dovich effect.  Rephaeli \& Lahav (1991) and Haehnelt \& Tegmark (1996) have discussed the possibility of the determination of the magnitude of the peculiar velocity by measuring the kinematic Sunyaev-Zel'dovich effect.  Observation of the kinematic Sunyaev-Zel'dovich effect has been done by Holzapfel et al. (1997c) and the limits on the peculiar velocities have been given for the clusters Abell 2163 and Abell 1689.  At the present stage the error bars are too large to give definite magnitudes of the peculiar velocities with the kinematic Sunyaev-Zel'dovich effect.  However, more accurate future observations are promising in the determination of the peculiar velocities.

  Therefore, relativistic corrections should be carefully taken into account in order to make the theoretical expressions for the kinematic Sunyaev-Zel'dovich effect as accurate as that for the thermal Sunyaev-Zel'dovich effect.  The purpose of the present paper is to calculate the relativistic correction to the kinematic Sunyaev-Zel'dovich effect.  In the present case, there exist two kinds of the relativistic corrections: one is due to the relativisitc electron temperature parameterized by $\theta_{e}$, and the other is due to the relativistic peculiar velocity parameterized by $\beta \equiv v/c$.  In particular we are interested in the interference contributions such as $O(\beta \theta_{e})$, $O(\beta \theta_{e}^{2})$ and $O(\beta^{2})$.  In the present paper we will derive the analytic expressions for the thermal and kinematic Sunyaev-Zel'dovich effects from a generalized Kompaneets equation by applying the Lorentz boost to the direction of the peculiar velocity of the cluster of galaxies.

  The present paper is organized as follows.  In $\S$ 2, the general formalism will be presented in deriving the thermal and kinematic Sunyaev-Zel'dovich effects on the same footing.  With the power series expansion approximation, an analytic expression including the relativistic effects will be derived for the thermal and kinematic Sunyaev-Zel'dovich effects.  Numerical results will be presented in $\S$ 3.  Finally  concluding remarks will be given in $\S$ 4.

\section{LORENTZ BOOSTED KOMPANEETS EQUATION}

  In the present section we will extend the Kompaneets equation to a system (the cluster of galaxies) moving with a peculiar velocity with respect to the CMBR.  We will formulate the kinetic equation for the photon distribution function using a relativistically covariant formalism (Berestetskii, Lifshitz, \& Pitaevskii 1982; Buchler \& Yueh 1976).  As a reference system, we choose the system which is fixed to the cosmic  microwave background radiation (CMBR).  The $z$-axis is fixed to a line connecting the observer and the center of mass of the cluster of galaxies (CG). (We assume that the observer is fixed to the CMBR frame.)  We fix the positive direction of the $z$-axis as the direction of the propagation of a photon from the cluster to the observer.  In this reference system, the center of mass of the cluster of galaxies is moving with a peculiar velocity $\vec{\beta} (\equiv \vec{v}/c$) with respect to the CMBR.  For simplicity, we choose the direction of the velocity in the $x$-$z$ plane, i.e. $\vec{\beta} = (\beta_{x}, 0, \beta_{z})$.

In the CMBR frame, the time evolution of the photon distribution 
function $n(\omega)$ is written as follows:
\begin{eqnarray}
\frac{\partial n(\omega)}{\partial t} & = & -2 \int \frac{d^{3}p}{(2\pi)^{3}} d^{3}p^{\prime} d^{3}k^{\prime} \, W \,
\left\{ n(\omega)[1 + n(\omega^{\prime})] f(E) - n(\omega^{\prime})[1 + n(\omega)] f(E^{\prime}) \right\} \, ,  \\
W & = & \frac{(e^{2}/4\pi)^{2} \, \overline{X} \, \delta^{4}(p+k-p^{\prime}-k^{\prime})}{2 \omega \omega^{\prime} E E^{\prime}} \, ,  \\
\overline{X} & = & - \left( \frac{\kappa}{\kappa^{\prime}} + \frac{\kappa^{\prime}}{\kappa} \right) + 4 m^{4} \left( \frac{1}{\kappa} + \frac{1}{\kappa^{\prime}} \right)^{2} 
 - 4 m^{2} \left( \frac{1}{\kappa} + \frac{1}{\kappa^{\prime}} \right) \, ,  \\
\kappa & = & - 2 (p \cdot k) \, = \, - 2 \omega E \left( 1 - \frac{\mid \vec{p} \mid}{E} {\rm cos} \alpha \right) \, ,  \\
\kappa^{\prime} & = &  2 (p \cdot k^{\prime}) \, = \, 2 \omega^{\prime} E \left( 1 - \frac{\mid \vec{p} \mid}{E} {\rm cos} \alpha^{\prime} \right) \, .
\end{eqnarray}
In the above $W$ is the transition probability corresponding to the Compton scattering.  The four-momenta of the initial electron and photon are $p = (E, \vec{p})$ and $k = (\omega, 0, 0, k)$, respectively.  The four-momenta of the final electron and photon are $p^{\prime} = (E^{\prime}, \vec{p}^{\prime})$ and $k^{\prime} = (\omega^{\prime}, \vec{k}^{\prime})$, respectively.  The angles $\alpha$ and $\alpha^{\prime}$ are the angles between $\vec{p}$ and $\vec{k}$, and between $\vec{p}$ and $\vec{k}^{\prime}$, respectively.  Throughout this paper, we use the natural unit $\hbar = c = 1$ unit, unless otherwise stated explicitly.  Here we note that eq.\ (2.1) assumes an isotropic photon distribution.  The kinematic effects lead to anisotropy but to a good approximation eq.\ (2.1) may be applied all the way through the cluster since the levels of anisotropy induced by the kinematic effect are small.

  The electron distribution functions in the initial and final states are Fermi--like in the CG frame.  They are related to the electron distribution functions in the CMBR frame as follows (Landau \& Lifshitz 1975):
\begin{eqnarray}
f(E) & = &  f_{C}(E_{C})  \, , \\
f(E^{\prime}) & = &  f_{C}(E_{C}^{\prime})  \,  ,  \\
E_{C} & = & \gamma \, \left(E - \vec{\beta} \cdot  \vec{p} \right) \, ,   \\
E_{C}^{\prime} & = & \gamma \, \left(E^{\prime} - \vec{\beta} \cdot \vec{p}^{\prime} \right) \, ,  \\
\gamma & \equiv & \frac{1}{\sqrt{1 - \beta^{2}}}   \, ,
\end{eqnarray}
where the suffix $C$ denotes the CG frame.  We caution the reader that the electron distribution function is anitotropic in the CMBR frame for $\beta \neq 0$.  By ignoring the degeneracy effects, we have the relativistic Maxwellian distribution for electrons with temperature $T_{e}$ as follows:
\begin{eqnarray}
f_{C}(E_{C}) & = & \left[ e^{\left\{(E_{C} - m)-(\mu_{C} - m) \right\}/k_{B}T_{e}} \, + \, 1 \right]^{-1}  \nonumber \\
& \approx & e^{-\left\{(E_{C}-m)-(\mu_{C} - m)\right\}/k_{B}T_{e}} \, ,
\end{eqnarray}
where $(\mu_{C} - m)$ is the non-relativistic chemical potential of the electron measured in the CG frame.  We now introduce the quantities
\begin{eqnarray}
x &  \equiv &  \frac{\omega}{k_{B}T_{e}}  \, ,  \\
\Delta x &  \equiv &  \frac{\omega^{\prime} - \omega}{k_{B}T_{e}}  \, .
\end{eqnarray}
Substituting eqs.\ (2.6) -- (2.13) into eq.\ (2.1), we obtain
\begin{eqnarray}
\frac{\partial n(\omega)}{\partial t} = -2 \int \frac{d^{3}p}{(2\pi)^{3}} d^{3}p^{\prime} d^{3}k^{\prime} \, W \, f_{C}(E_{C}) \,
\left[ \, \left\{ \, 1 + n(\omega^{\prime}) \, \right\} n(\omega) \,
  \right.  \hspace{3.0cm}  \nonumber  \\
 \left. \, - \,  \left\{ \, 1 + n(\omega) \, \right\} n(\omega^{\prime}) \, {\rm e}^{ \Delta x \gamma (1 - \vec{\beta} \cdot \hat{k}^{\prime} ) } \, {\rm e}^{ x \gamma \vec{\beta} \cdot ( \hat{k} - \hat{k}^{\prime} ) } \right] \, ,
\end{eqnarray}
where $\hat{k}$ and $\hat{k}^{ \prime}$ are the unit vectors in the directions of $\vec{k}$ and $\vec{k}^{ \prime}$, respectively.  Eq.\ (2.14) is our basic equation.

   We now expand eq.\ (2.14) in powers of 
$\Delta x$ by assuming $\Delta x  \, \ll 1$.  We obtain the Fokker-Planck expansion
\begin{eqnarray}
\frac{ \partial n(\omega)}{ \partial t} & = & 
 2 \left[ \frac{ \partial n}{ \partial x} \, I_{1,0} + n(1+n) \, I_{1,1} \right]
  \nonumber  \\
& + & 2 \left[ \frac{ \partial^{2} n}{ \partial x^{2}} \, I_{2,0}
+ 2(1+n) \frac{ \partial n}{ \partial x} \, I_{2,1} + n(1+n)  \, I_{2,2} \right]
  \nonumber  \\
& + & 2 \left[\frac{ \partial^{3} n}{ \partial x^{3}} \, I_{3,0}
+ 3(1+n) \frac{ \partial^{2} n}{ \partial x^{2}} \, I_{3,1}
+ 3(1+n) \frac{ \partial n}{ \partial x} \, I_{3,2} + n(1+n) \, I_{3,3} \right]
  \nonumber \\
& + & \cdot \cdot \cdot  \,  \nonumber \\
& + & 2 \, n \, \left[ (1 + n) J_{0} + \frac{ \partial n}{ \partial x } \,  J_{1} + \frac{ \partial^{2} n}{ \partial x^{2}} \, J_{2} + \frac{ \partial^{3} n}{ \partial x^{3}} \, J_{3} + \cdot \cdot \cdot  \, \, \,  \right] \, \, ,
\end{eqnarray}
where
\begin{eqnarray}
I_{k, \ell} & \equiv & \frac{1}{k !} \int \frac{d^{3}p}{(2\pi)^{3}} d^{3}p^{\prime} d^{3}k^{\prime} \, W \, f_{C}(E_{C}) \, (\Delta x)^{k}  \,  {\rm e}^{ x \gamma \vec{\beta} \cdot ( \hat{k} - \hat{k}^{\prime} ) }  
\gamma^{ \ell} \left( 1 - \vec{\beta} \cdot \hat{k}^{\prime} 
\right)^{ \ell}  \, , \\
  \nonumber \\
J_{k} & \equiv & \frac{-1}{k !} \int \frac{d^{3}p}{(2\pi)^{3}} d^{3}p^{\prime} d^{3}k^{\prime} \, W \, f_{C}(E_{C}) \, (\Delta x)^{k}  \left( \, 1 \, - \, {\rm e}^{ x \gamma \vec{\beta} \cdot ( \hat{k} - \hat{k}^{\prime} ) }  \right)  \, .
\end{eqnarray}
Analytic integration of eqs.\ (2.16) and (2.17) can be done with the power series expansion approximation of the integrand in terms of the electron momentum $p$.  In Itoh, Kohyama \& Nozawa (1998), the systematic analysis has been done in order to examine the accuracy of the power series expansion approximation.  It has been found that the power series expansion approximation is sufficiently accurate for $k_{B}T_{e} \leq 15$keV by taking into account $O(\theta_{e}^{5})$ corrections.

In addition to $\theta_{e}$, there is another parameter $\vec{\beta}$ in eqs.\ (2.16) and (2.17).  For most of the cluster of galaxies, $\beta \ll 1$ is realized.  For example, $\beta \approx $ 1/300 for a typical value of the peculiar velocity $v$=1,000km/s.  Therefore it should be sufficient to expand eqs. (2.16) and (2.17) in powers of $\beta$ and to retain up to  $O(\beta^{2})$ contributions.  We assume the initial photon distribution of the CMBR to be Planckian with a temperature $T_{0}$:
\begin{equation}
n_{0} (X) \, = \, \frac{1}{e^{X} - 1} \, , 
\end{equation}
where
\begin{equation}
X \, \equiv \, \frac{\omega}{k_{B} T_{0}}  \, .
\end{equation}

Substituting the results of eqs.\ (2.16) and (2.17) into eq.\ (2.15) and assuming $T_{0}/T_{e} \ll 1$, one obtains the following expression for the fractional distortion of the photon spectrum:

\begin{eqnarray}
\frac{\Delta n(X)}{n_{0}(X)} & = & \frac{y \, X e^{X}}{e^{X}-1} \, \theta_{e} \, \left[  \, \,
Y_{0} \, + \, \theta_{e} Y_{1} \, + \, \theta_{e}^{2} Y_{2} \, + \,  \theta_{e}^{3} Y_{3} \, + \, \theta_{e}^{4} Y_{4} \,  \right]  \,   \nonumber  \\
  & + & \frac{y \, X e^{X}}{e^{X}-1}  \, \beta^{2} \, \left[ \, \, \frac{1}{3} Y_{0} \, + \, \theta_{e} \left( \, \frac{5}{6} Y_{0} \, + \, \frac{2}{3} Y_{1} \, \right) \, \right]    \, \nonumber \\ 
  & + & \frac{y \, X e^{X}}{e^{X}-1}  \, \beta \, P_{1}(\hat{\beta}_{z}) \, \left[ \, \, 1 \, + \, \theta_{e} C_{1} \, + \, \theta_{e}^{2} C_{2} \,  \right]    \, \nonumber \\
& + & \frac{y \, X e^{X}}{e^{X}-1} \beta^{2}  P_{2} (\hat{\beta}_{z}) \, \left[ \, D_{0} \, + \, \theta_{e} D_{1} \, \right] \, ,  \\
 \nonumber \\
\hat{\beta}_{z} & \equiv & \frac{\beta_{z}}{\beta} \, = \, {\rm cos} \theta_{\gamma}  \, , \\
P_{1} (\hat{\beta}_{z}) & = & \hat{\beta}_{z} \, , \\
P_{2} (\hat{\beta}_{z}) & = & \frac{1}{2} \, \left( 3 \hat{\beta}_{z}^{2} - 1 \right) \, ,
\end{eqnarray}
where $\theta_{\gamma}$ is the angle between the directions of the peculiar velocity of the cluster ($\vec{\beta}$) and the initial photon momentum ($\vec{k}$) which is chosen as the positive $z$-direction.  The reader should remark that this sign convention for the positive $z$-direction is opposite to the ordinary one.  Thus a cluster moving away from the observer has $\hat{\beta}_{z} < 0$.  However, because of the positive sign in front of the $P_{1}(\hat{\beta}_{z})$ term, one obtains $\Delta n(X) < 0$ in this case, as one should.  The coefficients are defined as follows:
\begin{eqnarray}
Y_{0} & = & - 4 \, + \tilde{X}  \,  , \\
Y_{1} & = & - 10 + \frac{47}{2} \tilde{X} - \frac{42}{5} \tilde{X}^{2} + \frac{7}{10} \tilde{X}^{3}  \, + \, \tilde{S}^{2} \left( - \frac{21}{5} + \frac{7}{5} \tilde{X} \right) \,  ,  \\
Y_{2} & = & - \frac{15}{2} + \frac{1023}{8} \tilde{X} - \frac{868}{5} \tilde{X}^{2} + \frac{329}{5} \tilde{X}^{3} - \frac{44}{5} \tilde{X}^{4} + \frac{11}{30} \tilde{X}^{5}  \nonumber \\ 
& & + \tilde{S}^{2} \left( - \frac{434}{5} + \frac{658}{5} \tilde{X}  - \frac{242}{5}  \tilde{X}^{2} + \frac{143}{30} \tilde{X}^{3} \right) 
 +  \tilde{S}^{4} \left( - \frac{44}{5} + \frac{187}{60} \tilde{X} \right) \, ,   \\
Y_{3} & = & \frac{15}{2} + \frac{2505}{8} \tilde{X} - \frac{7098}{5} \tilde{X}^{2} + \frac{14253}{10} \tilde{X}^{3} - \frac{18594}{35} \tilde{X}^{4}   \nonumber  \\
& + &  \frac{12059}{140} \tilde{X}^{5} - \frac{128}{21} \tilde{X}^{6} + \frac{16}{105} \tilde{X}^{7} \nonumber \\ 
& + & \tilde{S}^{2} \left( - \frac{7098}{10} + \frac{14253}{5} \tilde{X} - \frac{102267}{35}  \tilde{X}^{2} + \frac{156767}{140} \tilde{X}^{3} - \frac{1216}{7}  \tilde{X}^{4} + \frac{64}{7} \tilde{X}^{5} \right)  \nonumber  \\
& + &  \tilde{S}^{4} \left( - \frac{18594}{35} + \frac{205003}{280} \tilde{X} - \frac{1920}{7}  \tilde{X}^{2} + \frac{1024}{35} \tilde{X}^{3} \right) \nonumber  \\
& + &  \tilde{S}^{6} \left( - \frac{544}{21} + \frac{992}{105} \tilde{X} \right) \, , \\
Y_{4} & = & - \frac{135}{32} + \frac{30375}{128} \tilde{X} - \frac{62391}{10} \tilde{X}^{2} + \frac{614727}{40} \tilde{X}^{3} - \frac{124389}{10} \tilde{X}^{4}   \nonumber  \\
& + &  \frac{355703}{80} \tilde{X}^{5} - \frac{16568}{21} \tilde{X}^{6} + \frac{7516}{105} \tilde{X}^{7} - \frac{22}{7} \tilde{X}^{8} + \frac{11}{210} \tilde{X}^{9} \nonumber \\ 
& + & \tilde{S}^{2} \left( - \frac{62391}{20} + \frac{614727}{20} \tilde{X} - \frac{1368279}{20} \tilde{X}^{2} + \frac{4624139}{80} \tilde{X}^{3} - \frac{157396}{7}  \tilde{X}^{4}  \right. \nonumber  \\
&  & \, \, \, \, \, + \, \left. \frac{30064}{7} \tilde{X}^{5} - \frac{2717}{7} \tilde{X}^{6} + \frac{2761}{210} \tilde{X}^{7}   \right)  \nonumber  \\
& + &  \tilde{S}^{4} \left( - \frac{124389}{10} + \frac{6046951}{160} \tilde{X} - \frac{248520}{7} \tilde{X}^{2} + \frac{481024}{35} \tilde{X}^{3} - \frac{15972}{7} \tilde{X}^{4}  \right. \nonumber  \\
&  &  \, \, \, \, + \, \left. \frac{18689}{140} \tilde{X}^{5}  \right) \nonumber  \\
& + &  \tilde{S}^{6} \left( - \frac{70414}{21} + \frac{465992}{105} \tilde{X} - \frac{11792}{7} \tilde{X}^{2} + \frac{19778}{105} \tilde{X}^{3} \right) \nonumber  \\
& + &  \tilde{S}^{8} \left( - \frac{682}{7} + \frac{7601}{210} \tilde{X} \right) \, , \\
\nonumber  \\
C_{1} & = & 10 - \frac{47}{5} \tilde{X} + \frac{7}{5} \tilde{X}^{2} + \frac{7}{10} \tilde{S}^{2}  \,  ,  \\
C_{2} & = & 25 - \frac{1117}{10} \tilde{X} + \frac{847}{10} \tilde{X}^{2} - \frac{183}{10} \tilde{X}^{3} + \frac{11}{10} \tilde{X}^{4}    \nonumber \\ 
& & + \tilde{S}^{2} \left( \frac{847}{20} - \frac{183}{5} \tilde{X}  + \frac{121}{20}  \tilde{X}^{2} \right)  +  \frac{11}{10} \tilde{S}^{4}  \, ,   \\
\nonumber \\
D_{0} & = & - \frac{2}{3} + \frac{11}{30} \tilde{X} \, , \\
D_{1} & = & - 4 + 12 \tilde{X} - 6 \tilde{X}^{2} + \frac{19}{30} \tilde{X}^{3}  \, + \, \tilde{S}^{2} \left( - 3 + \frac{19}{15} \tilde{X} \right) \,  ,
\end{eqnarray}
and
\begin{eqnarray}
y & \equiv & \sigma_{T} \int d \ell N_{e}  \, , \\
\tilde{X} & \equiv &  X \, {\rm coth} \left( \frac{X}{2} \right)  \, , \\
\tilde{S} & \equiv & \frac{X}{ \displaystyle{ {\rm sinh} \left( \frac{X}{2} \right)} }   \, ,
\end{eqnarray}
where $N_{e}$ is the electron number density in the CG frame, the integral in eq.\ (2.33) is over the photon path length in the cluster, and $\sigma_{T}$ is the Thomson cross section.

Eq.\ (2.20) is one of our main results.  The first line in eq.\ (2.20) is the expression of the thermal Sunyaev-Zel'dovich effect.  The relativistic corrections up to $O(\theta^{3}_{e})$ was first derived by Challinor \& Lasenby (1998) and also derived by Itoh, Kohyama \& Nozawa (1998) for up to $O(\theta_{e}^{5})$ corrections.  The second line corresponds to the $O(\beta^{2})$ corrections to the thermal Sunyaev-Zel'dovich effect.  The third and fourth lines correspond to the kinematic Sunyaev-Zel'dovich effects of the first and second orders in $\beta$, respectively.  One notes that the first term in the third line of eq.\ (2.20) corresponds to the lowest-order kinematic Sunyaev-Zel'dovich effect (Sunyaev \& Zel'dovich 1980b).  The following remarks should be emphasized.  In the present approach the kinematic Sunyaev-Zel'dovich effects have been derived in a straightforward manner from the standard equation for the thermal Sunyaev-Zel'dovich effect by applying the Lorentz boost to the initial and final electrons.

Concerning the photon number conservation of eq.\ (2.20), the following remarks should be also noted.  As for the thermal Sunyaev-Zel'dovich effect terms, the photon number conservation is satisfied as discussed in Itoh, Kohyama \& Nozawa (1998) as well as in Challinor \& Lasenby (1998).  Namely, $Y_{0}$, $Y_{1}$, $Y_{2}$, $Y_{3}$ and $Y_{4}$ terms separately vanish by the integration $\int dX X^{2}$.  Therefore the first and second lines vanish by the integration.  On the other hand, the third and fourth lines are propotional to $P_{1}(\hat{\beta}_{z})$ and  $P_{2}(\hat{\beta}_{z})$, respectively.  Therefore the third and fourth lines vanish by the integration over the solid angle $\int d \Omega_{\gamma}$.  Therefore the photon number conservation is guaranteed for eq.\ (2.20).

Finally we define the distortion of the spectral intensity as follows:
\begin{equation}
\Delta I \, = \, \frac{X^{3}}{e^{X}-1} \frac{\Delta n(X)}{n_{0}(X)} \, .
\end{equation}

\section{NUMERICAL ANALYSES}

  We now study the relativistic contribution for the kinematic Sunyaev-Zel'dovich effect.  In Fig.\ 1 we have plotted the distortion of the spectral intensity $\Delta I/y$ as a function of $X$ for a parameter set, $k_{B}T_{e}$=10keV, $\beta_{z}=1/300$.  The thermal and kinematic Sunyaev-Zel'dovich effects have been plotted.  For the kinematic Sunyaev-Zel'dovich effect, the absolute value of the effect is shown.  For a receding cluster one has $\Delta I < 0$, whereas for an approaching cluster one has $\Delta I > 0$.  We have shown the decomposition of the absolute value of the kinematic Sunyaev-Zel'dovich contribution in Fig.\ 2.  The dashed curve includes only the leading order $O(\beta)$ term.  The dotdashed curve includes up to the $O(\beta \theta_{e})$ terms.  The solid curve includes up to $O(\beta \theta_{e}^{2})$ terms.  The dotted curve is the contribution of the $O(\beta^{2})$ terms.  It is clear from Fig.\ 2 that the relativistic correction is extremely important, which is roughly 8\% of the leading order contribution of the kinematic Sunyaev-Zel'dovich effect.  It is also clear from Fig.\ 2 that the $O(\beta^{2})$ correction is very small and it can be safely neglected.

\subsection{RAYLEIGH--JEANS REGION}

In the Rayleigh--Jeans limit where $X \rightarrow 0$, eq.\ (2.20) is further simplified:
\begin{eqnarray}
\frac{\Delta n(X)}{n_{0}(X)} & \rightarrow & - 2 y \, \theta_{e} \, \left[ \, 1 - \frac{17}{10} \theta_{e} + \frac{123}{40} \theta_{e}^{2} - \frac{1989}{280} \theta_{e}^{3} + \frac{14403}{640} \theta_{e}^{4} \, \right]  \, \nonumber \\
 & & - 2 y \, \beta^{2} \left[ \, \frac{1}{3} \, - \, \frac{3}{10} \, \theta_{e} \, \right]  \,   \nonumber  \\
 & &  + \, \, y \, \beta \, P_{1}(\hat{\beta}_{z}) \left[ \, 1 -  \frac{2}{5} \theta_{e} + \frac{13}{5} \theta_{e}^{2} \, \right] \, \nonumber \\
 & & +  \, \, y \, \beta^{2} \, P_{2}( \hat{\beta}_{z}) \left[ \,  \frac{1}{15} \, - \, \frac{4}{5} \theta_{e} \, \right] \, .
\end{eqnarray}
For this expression also, one should note the sign convention for the positive $z$-axis stated below eq.\ (2.23).  For the purpose of illustration we have calculated the kinematic Sunyaev-Zel'dovich effect for a parameter set $k_{B} T_{e}$ = 10keV, $\beta_{z}$ = 1/300.  In this region, the leading order kinematic Sunyaev-Zel'dovich effect is 8.8\% of the thermal Sunyaev-Zel'dovich effect.  The relativistic corrections of the $O(\beta \theta_{e})$ and $O(\beta^{2})$ terms are 0.07\% and 0.02\% of the thermal Sunyaev-Zel'dovich effect, respectively.  Therefore the relativistic corrections of $O(\beta \theta_{e})$ and $O(\beta^{2})$ are safely neglected.

\subsection{CROSSOVER FREQUENCY REGION}

It is well known that the kinematic Sunyaev-Zel'dovich effect becomes extremely important in the crossover frequency region, where the thermal Sunyaev-Zel'dovich effect vanishes.  Therefore the accurate position of the crossover frequency $X_{0}$ is extremely important.  In Itoh, Kohyama \& Nozawa (1998), the following fitting formula for $X_{0}$ has been obtained.
\begin{equation}
X_{0} \, = \, 3.830 \, \left( \, 1 + \, 1.1674 \theta_{e} \, - \, 0.8533 \theta_{e}^{2} \, \right)  \, .
\end{equation}
The errors of this fitting function are less than $1 \times 10^{-3}$ for $0 \leq k_{B}T_{e} \leq 50{\rm keV}$.

Substituting eq.\ (3.2) into eq.\ (2.36), we have calculated the kinematic Sunyaev-Zel'dovich effect at $X$ = $X_{0}$ as a function of $k_{B} T_{e}$.  The result has been plotted in Fig.\ 3.  The dashed curve is the leading order $O(\beta)$ contribution.  The dotdashed curve includes up to the  $O(\beta \theta_{e})$ contribution.  The solid curve includes up to the  $O(\beta \theta_{e}^{2})$ contribution.  The $O(\beta^{2})$ contribution is very small and is again safely neglected.  It is clear that the relativistic corrections become significant as the electron temperature increases.  At a typical electron temperature $k_{B}T_{e}$ = 10keV, one obtains the $-8.2\%$ and $+1.3\%$ corrections from the $O(\beta \theta_{e})$ and $O(\beta \theta_{e}^{2})$ contributions, respectively.  The $O(\beta^{2})$ correction is extremely small.  It is $+0.2\%$ for $\beta=1/300$ at $k_{B} T_{e}$ = 10keV.  Therefore one can safely neglect the $O(\beta^{2})$ corrections.  Although the thermal Sunyaev-Zel'dovich effect is exactly zero at $X$ = $X_{0}$, the slope is very large as shown in Fig.\ 1.  At $k_{B}T_{e}$ = 10keV, for example, the thermal Sunyaev-Zel'dovich effect becomes $\pm$17\% of the kinematic Sunyaev-Zel'dovich effect if one shifts $X_{0}$ by $\pm$1\%.  Therefore the precise determination of the position of $X_{0}$ is extremely crucial in isolating the kinematic Sunyaev-Zel'dovich effect.

It is well known that the kinematic Sunyaev-Zel'dovich effect is useful in determining the peculiar velocity; see for example, Sunyaev \& Zel'dovich (1980b) and Rephaeli \& Lahav (1991).  The relativistic corrections are directly reflected on the determination of the peculiar velocity of the cluster of galaxies with the observation of the kinematic Sunyaev-Zel'dovich effect.  The crossover frequency is shifted toward higher (lower) frequency region by the kinematic Sunyaev-Zel'dovich contribution when the cluster is moving outward (inward) with respect to the observer.  The shift is given by $\Delta X_{0} \equiv \, X_{0} - X_{0,\beta}$, where $X_{0}$ is given by eq.\ (3.2) and $X_{0,\beta}$ is the new crossover frequency in the presence of the kinematic Sunyaev-Zel'dovich effect.  We define a function $h$ as follows.
\begin{equation}
h \, \equiv \, \Delta X_{0} / \left( 300 \beta \, P_{1}(\hat{\beta}_{z}) \right)  \, .
\end{equation}
In Fig.\ 4 we have plotted $h$ as a function of $k_{B} T_{e}$ for the typical peculiar velocities $v$ = 500km/s, 1,000km/s and 1,500km/s, which correspond to $\beta$ = 0.5/300, 1/300 and 1.5/300, respectively.  These curves show an excellent scaling for 5keV $\leq k_{B} T_{e} \leq$ 15keV.  The function $h$ can be fitted as follows.
\begin{eqnarray}
h & = & \frac{a_{1}}{\theta_{e} - \theta_{e,min}} \, + \, \frac{a_{2}}{ \left( \theta_{e} - \theta_{e,min} \right)^{2} } \, + \, a_{3} \, + \, a_{4} \, \theta_{e} \, + \, a_{5} \, \theta_{e}^{2}  \, , 
\end{eqnarray}
where $\theta_{e,min}$ is the minimum value of $\theta_{e}$ which has a real solution for a given $\beta$.  We fix $\theta_{e,min}$ = 1.654$\times 10^{-3}$.  The other constants are $a_{1}$ = 3.857$\times 10^{-3}$, $a_{2}$ = $-4.631 \times 10^{-6}$, $a_{3}$ = 1.370$\times 10^{-2}$, $a_{4}$ = 1.014$\times 10^{-2}$, and $a_{5}$ =  1.000$\times 10^{-2}$.  The errors of this fitting function are less than 4\% for $5 \leq k_{B}T_{e} \leq 15{\rm keV}$ and $0.5/300 \leq \beta \leq 1.5/300$.

\section{CONCLUDING REMARKS}

By applying the Lorentz boost to the standard formalism of the extended Kompaneets equation, we have derived a formalism of the kinematic Sunyaev-Zel'dovich effect for the cluster of galaxies with a peculiar velocity $\beta$.  With the power series expansion approximation in terms of the electron temperature $\theta_{e}$ and the peculiar velocity $\beta$, we have derived an analytic expression for the kinematic Sunyaev-Zel'dovich effect which includes the relativistic corrections of up to $O(\beta \theta_{e}^{2})$ and $O(\beta^{2} \theta_{e})$.  It has been found that the relativistic correction is significant.  At a typical temperature $k_{B}T_{e}$ = 10keV, one obtains the $-8.2\%$ and $+1.3\%$ corrections from the $O(\beta \theta_{e})$ and $O(\beta \theta_{e}^{2})$ contributions, respectively.  The $O(\beta^{2})$ correction is small.  It is $+0.2\%$ for $\beta=1/300$ at $k_{B}T_{e}$ = 10keV.  These relativistic corrections are directly reflected on the determination of the peculiar velocity of the cluster of galaxies with the observation of the kinematic Sunyaev-Zel'dovich effect.  Finally we have fitted the shift of the crossover frequency $\Delta X_{0}$.  The shift has an excellent scaling  for $5 \leq k_{B}T_{e} \leq 15{\rm keV}$ and $0.5/300 \leq \beta \leq 1.5/300$.  The errors of the fitting function are less than 4\%.

  We received a preprint ``Cosmic Microwave Background Radiation in the Direction of a Moving Cluster of Galaxies with Hot Gas: Relativistic Corrections" from S. Y. Sazonov and R. A. Sunyaev after the submittal of our original manuscript to The Astrophysical Journal.  Most of our original results agreed with theirs (their equation (12)) except for the term proportional to $\beta^{2}$ (the second term on the right-hand-side of equation (2.20) in the present paper).  Although this term makes a negligible contribution in the real situation of the clusters of galaxies, we endeavored to find out the cause for the discrepancy.  Later, this discrepancy was pointed out by our referee A. Challinor who carried out an independent calculation.  Finally, we have found out an error in our original calculation and have corrected it.  Now our result agrees with that of Sazonov and Sunyaev (although they stop at a lower order of expansion) as well as that of Challinor.  We sincerely wish to thank S. Y. Sazonov, R. A. Sunyaev, and A. Challinor for motivating us to obtain the correct term proportional to $\beta^{2}$.  We also wish to thank R. A. Sunyaev for his very informative communication to us.  Last but not least, we wish to thank A. Challinor for his invaluable work as our referee, examining our mathematical results thoroughly and giving us many useful suggestions.

\newpage


\references{} 
\reference{} Arnaud, K. A., Mushotzky, R. F., Ezawa, H., Fukazawa, Y., Ohashi, T., Bautz, M. W., Crewe, G. B., Gendreau, K. C., Yamashita, K., Kamata, Y., \& Akimoto, F. 1994, ApJ, 436, L67
\reference{} Berestetskii, V. B., Lifshitz, E. M., \& Pitaevskii, L. P. 1982, $Quantum$ $Electrodynamics$ (Oxford: Pergamon)
\reference{} Birkinshaw, M. 1979, MNRAS, 187, 847
\reference{} Birkinshaw, M., \& Hughes, J., P. 1994, ApJ, 420, 33
\reference{} Birkinshaw, M., Hughes, J. P., \& Arnaud, K. A. 1991, ApJ, 379, 466
\reference{} Buchler, J. R., \& Yueh, W. R. 1976, ApJ, 210, 440
\reference{} Cavaliere, A., Danese, L., \& De Zotti, G. 1979, A\&A, 75, 322
\reference{} Challinor, A., \& Lasenby, A. 1998, ApJ, in press
\reference{} Chandrasekhar, S. 1950, $Radiative$ $Transfer$ (New York: Dover)
\reference{} Gunn, J. E. 1978, in Observational Cosmology, 1, ed. A. Maeder, L. Martinet \& G. Tammann (Sauverny: Geneva Obs.)
\reference{} Herbig, T., Lawrence, C. R., Readhead, A. C. S., \& Gulkus, S. 1995, ApJ, 449, L5
\reference{} Haehnelt, M. G., \& Tegmark, M. 1996, Mon. Not. Roy. Astron. Soc. 279, 545
\reference{} Holzapfel, W. L. et al. 1997a, ApJ, 479, 17
\reference{} Holzapfel, W. L. et al. 1997b, ApJ, 480, 449
\reference{} Holzapfel, W. L. et al. 1997c, ApJ, 481, 35
\reference{} Itoh, N., Kohyama, Y. , \& Nozawa, S. (1998), ApJ, in press
\reference{} Jones, M. 1995, Astrophys. Lett. Comm., 6, 347
\reference{} Kompaneets, A. S. 1957, Soviet Physics JETP, 4, 730
\reference{} Landau, L. D., \& Lifshitz, E. M. 1975, $The$ $Classical$ $Theory$ $of$ $Fields$ (Oxford: Pergamon) 
\reference{} Markevitch, M., Mushotzky, R., Inoue, H., Yamashita, K., Furuzawa, A., \& Tawara, Y. 1996, ApJ, 456, 437
\reference{} Markevitch, M., Yamashita, K., Furuzawa, A., \& Tawara, Y. 1994, ApJ, 436, L71
\reference{} Myers, S. T., Baker, J. E., Readhead, A. C. S., \& Herbig, T. 1995, preprint
\reference{} Rephaeli, Y. 1995, ApJ, 445, 33
\reference{} Rephaeli, Y., \& Lahav, O. 1991, ApJ, 372, 21
\reference{} Rephaeli. Y., \& Yankovitch, D. 1997, ApJ, 481, L55
\reference{} Sazonov, S. Y., \& Sunyaev, R. A. 1998, preprint astro-ph/9804125 v2
\reference{} Silk, J. I., \& White, S. D. M. 1978, ApJ, 226, L103
\reference{} Stebbins, A., 1997, preprint astro-ph/9705178
\reference{} Sunyaev, R. A., \& Zel'dovich, Ya. B. 1980a, Ann. Rev. Astron. Astrophys., 18, 537
\reference{} Sunyaev, R. A., \& Zel'dovich, Ya. B. 1980b, Mon. Not. R. astro. Soc., 190, 413
\reference{} Sunyaev, R. A., \& Zel'dovich, Ya. B. 1972, Comm. Ap. Space Sci., 4, 173
\reference{} Weymann, R. 1965, Phys. Fluid, 8, 2112
\reference{} Zel'dovich, Ya. B., \& Sunyaev, R. A. 1969, Astrophys. Space Sci., 4, 301


\newpage

\centerline{\bf \large Figure Captions}

\begin{itemize}

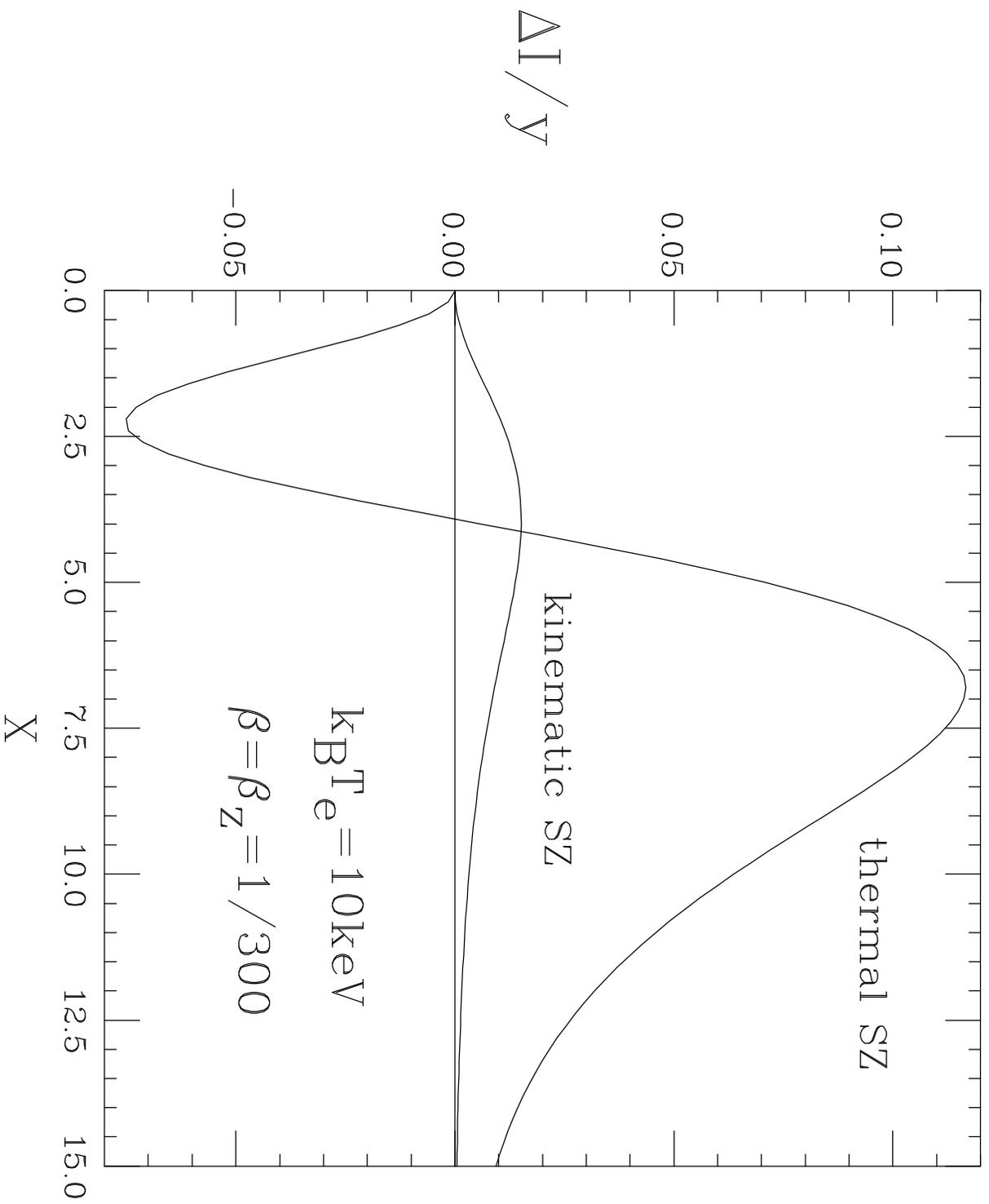
\item Fig.1. Spectral intensity distortion $\Delta I/y$ as a function of $X$ for $k_{B}T_{e}$ = 10keV, $\beta=\beta_{z}$=1/300.  The thermal and kinematic Sunyaev-Zeldovich effects have been plotted.  For the kinematic Sunyaev-Zel'dovich effect, the absolute value of the effect is shown.

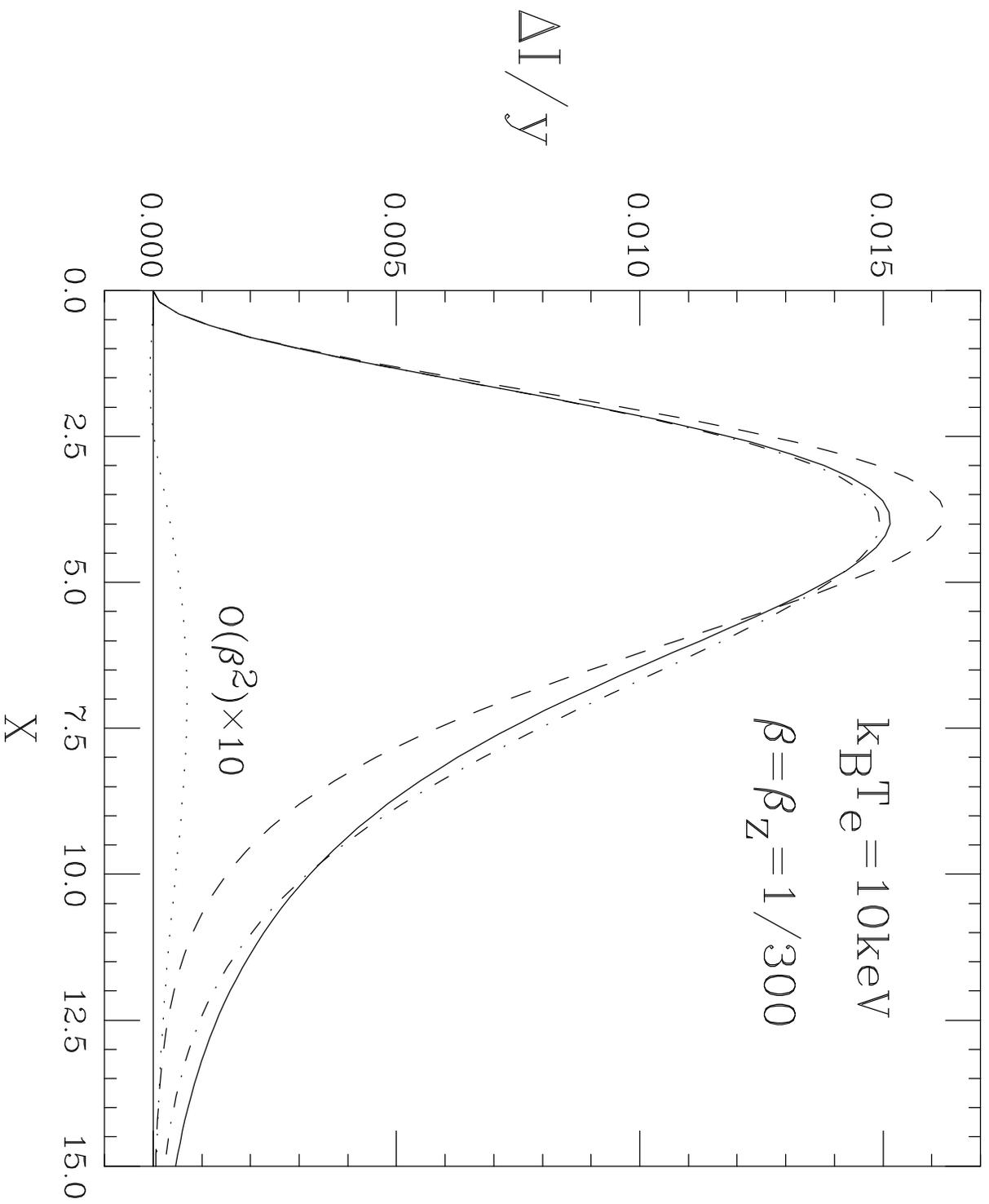
\item Fig.2. The decomposition of the absolute value of the kinematic Sunyaev-Zel'dovich contribution for $k_{B}T_{e}$ = 10keV, $\beta=\beta_{z}$=1/300.  The dashed curve is the leading order $O(\beta)$ term.  The dotdashed curve is the $O(\beta)+O(\beta \theta_{e})$ terms.  The solid curve is the $O(\beta)+O(\beta \theta_{e})+O(\beta \theta_{e}^{2})$ terms.  The dotted curve is the contribution of the $O(\beta^{2})$ terms, where a factor 10 was multiplied in order to make the curve visible in this figure.

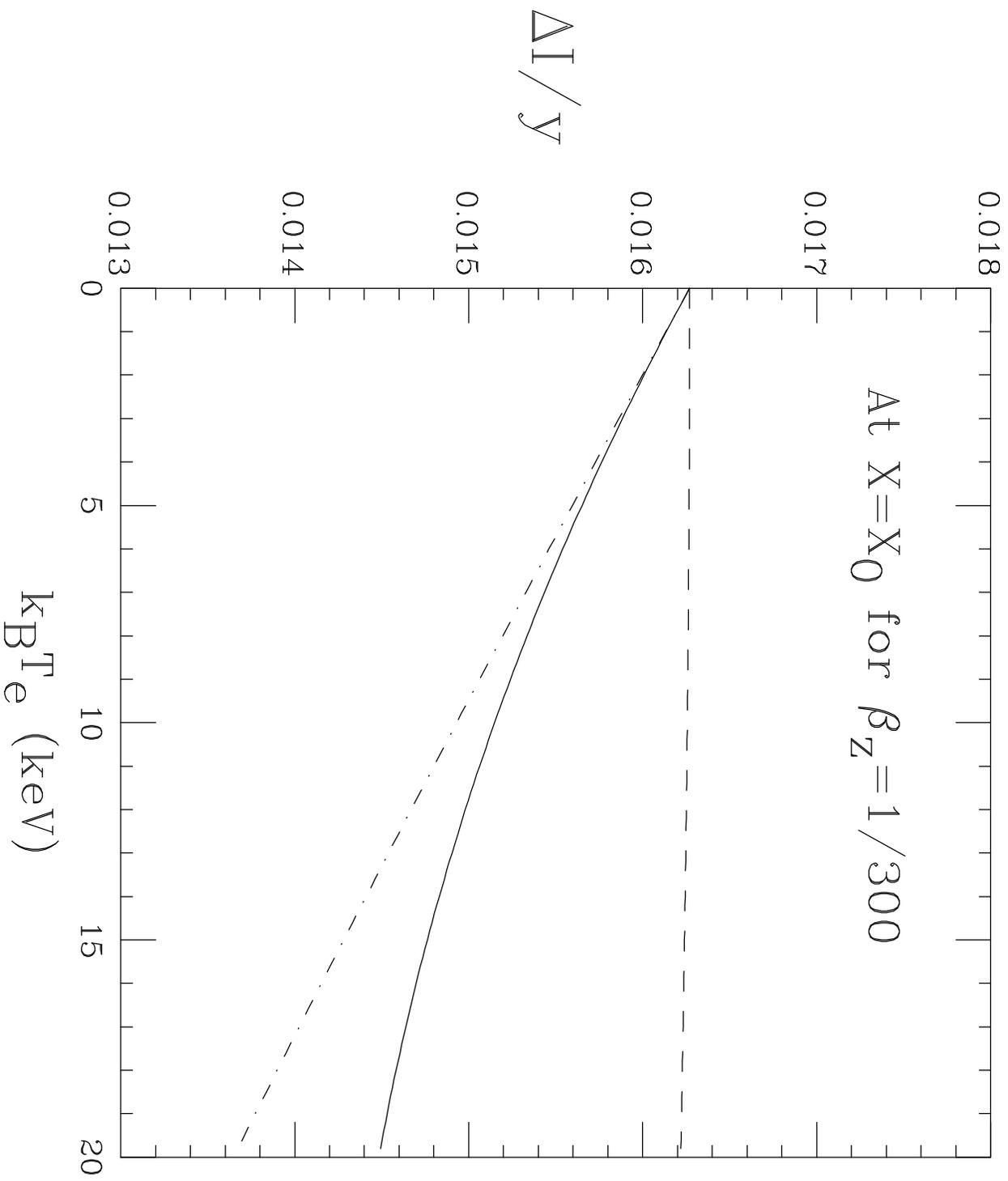
\item Fig.3. The absolute value of the kinematic Sunyaev-Zel'dovich effect at $X$ = $X_{0}$ as a function of $k_{B} T_{e}$ for $\beta_{z}$=1/300.  The dashed curve is the leading order $O(\beta)$ term.  The dotdashed curve is the $O(\beta)+O(\beta \theta_{e})$ terms.  The solid curve is the $O(\beta)+O(\beta \theta_{e})+O(\beta \theta_{e}^{2})$ terms.

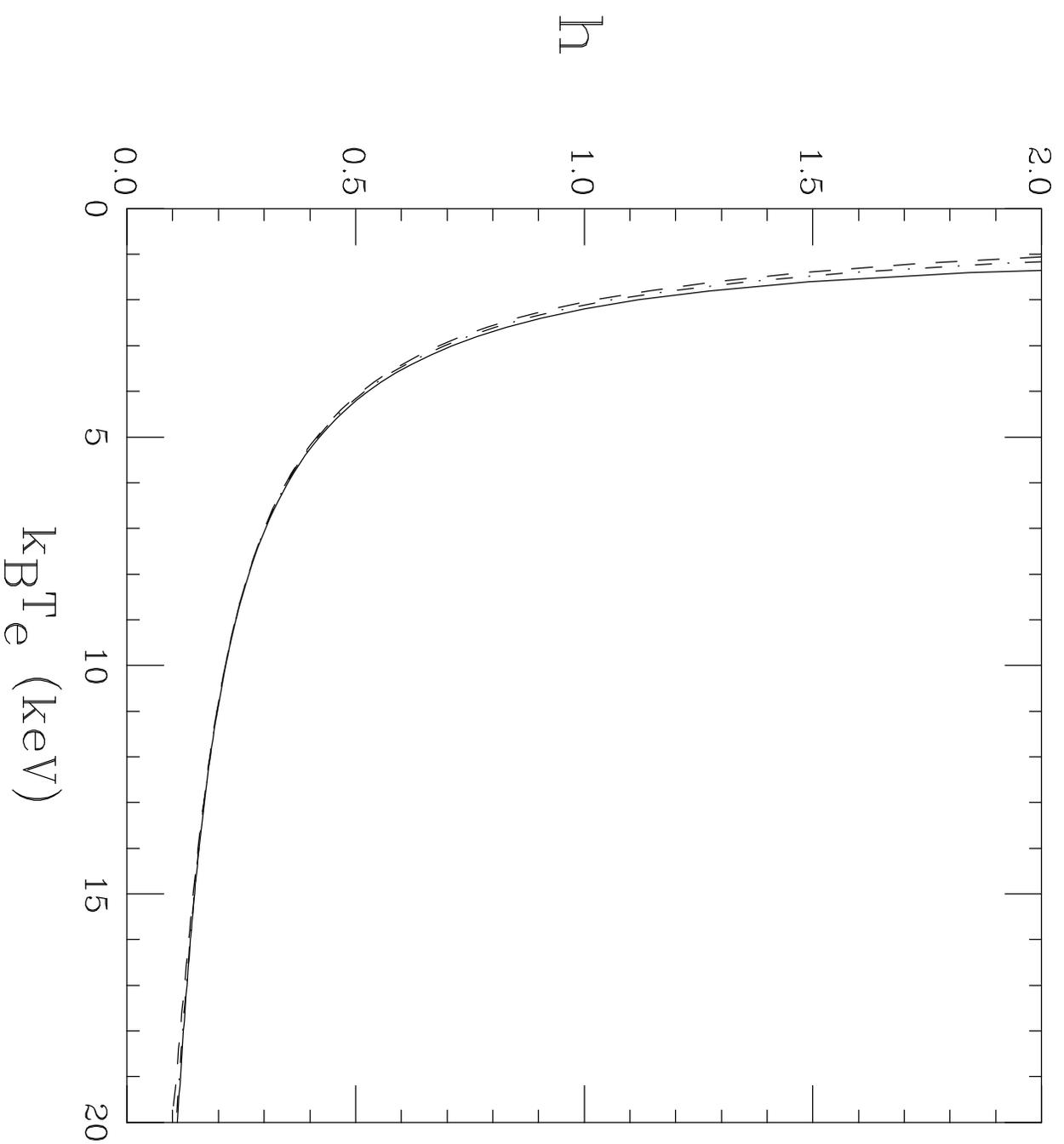
\item Fig.4. The shift $h$ of the crossover frequency as a function of $k_{B} T_{e}$.  The dashed, dotdashed and solid curves correspond to $v$ = 500km/s, 1,000km/s and 1,500km/s, respectively.

\end{itemize}

\end{document}